\documentclass[structabstract]{aa}  
\usepackage{graphicx}
\usepackage{color}
\usepackage{subfigure}
\usepackage{url}
\usepackage[authoryear]{natbib}
\usepackage{txfonts}
\begin{document}
   \title{SPS: A software simulator for the Herschel-SPIRE photometer}
%\titlerunning{The SPIRE Photometer Simulator}

%   \subtitle{I. Overviewing the $\kappa$-mechanism}

   \author{B. Sibthorpe\inst{1}
          \and
					 P. Chanial\inst{2}
          \and
          M. J. Griffin\inst{3}%\fnmsep
          }

   \institute{UK Astronomy Technology Centre, Royal Observatory Edinburgh, Blackford Hill, Edinburgh, EH9 3HJ\\
              \email{sib@roe.ac.uk}
         \and
             Astrophysics Group, Blackett Laboratory, Imperial College, Prince Consort Road, London SW7 2AZ, UK\\
             \email{p.chanial@imperial.ac.uk}
         \and
             Cardiff School of Physics and Astronomy, Cardiff University, Queens Buildings, The Parade, Cardiff, CF24 3AA\\
             \email{matt.griffin@astro.cf.ac.uk}
             }

   \date{Received XXXX XX, 2008; accepted XXXX XX, 2008}

% \abstract{}{}{}{}{} 
% 5 {} token are mandatory
 
  \abstract
   {}
   {Instrument simulators are becoming ever more useful for planning and analysing large astronomy survey data. In this paper we present a simulator for the Herschel-SPIRE photometer. We describe the models it uses and the form of the input and output data.}
   {The SPIRE photometer simulator is a software package which uses theoretical models, along with flight model test data, to perform numerical  simulations of the output time-lines from the instrument in operation on board the Herschel space observatory.}
   {A description of the types of uses of the simulator are given, along with information on its past uses. These include example simulations performed in preparation for a high redshift galaxy survey, and a debris disc survey. These are presented as a demonstration of the sort of outputs the simulator is capable of producing.}
   {}


   \maketitle
%
%________________________________________________________________________________________________________________________________
\section{Introduction}
Software simulators are now used in a wide range of astronomy instrument projects. They have many uses, from guiding the instrument design, through observation planning, and finally helping to understand the astronomical data. This paper describes the software package developed to simulate the Spectral and Photometric Imaging Receiver, SPIRE \citep{Griffin2008} photometer instrument on board the European Space Agency's Herschel space observatory \citep{Pilbratt2008}, scheduled for launch in 2009. This software is designed to simulate the behavior of the instrument and spacecraft, and return realistic data in a format compatible with the Herschel data pipeline \citep{Griffin2008_pipeline}.

SPIRE is a dual instrument, comprising a three-band submillimetre camera and an imaging Fourier transform spectrometer (FTS). In this paper we will only be considering the photometer; for information on the spectrometer simulator see \cite{Lindner}. The SPIRE photometer uses arrays of hexagonally packed feedhorn-coupled bolometric detectors, and has a field of view (FoV) of 4 x 8 arcminutes. The FoV is observed simultaneously in three bands centred approximately at 250, 350 and 500 $\mu$m, giving diffraction limited full-width-half-maximum (FWHM) beam widths of approximately 18, 25 and 36 arcseconds respectively. There are 139, 88 and 43 detectors in the 250 $\mu$m, 350 $\mu$m and 500 $\mu$m arrays respectively, along with two `dark pixels' -- bolometers positioned outside the instrument FoV -- and two thermistors located on each array. 

Maximising the aperture efficiency of the feedhorns requires an aperture corresponding to an angle of $2\lambda/D$ on the sky, where $\lambda$ is the wavelength and $D$ is the telescope diameter. Consequently, the detector beams have an angular separation of approximately twice the FWHM beam size on the sky. As a result, specific observing patterns -- either jiggling or scanning -- must be employed to achieve Nyquist sampling of the sky brightness distribution \citep{GBG}. Jiggling is achieved using a dual axis internal beam steering mirror (BSM), while fully sampled scanned observations are obtained by scanning the telescope at the specific scanning angle of 42.4$^\circ$ with respect to short axis of the bolometer array \citep{Sibthorpe2006, Waskett2007}.

The bolometer signal time-lines are filtered on-board by a 5-Hz low-pass filter. After multiplexing, an appropriate 4-bit offset is removed from each data time-line before the final gain stage, after which the data are digitized and telemetered to the ground.  The offset removal allows the data to be sampled with 20 bit accuracy using a 16-bit ADC.  The timelines are reconstructed to 20-bit accuracy on the ground using the telemtered offset and digitized data time-lines. The photometer on-board signal chain is described in detail by \cite{Griffin2008_pipeline}.

In Section 2 of this paper we describe the simulation software itself, and its operation. Details of the format of the simulator input and output data are also given. Section 3 presents some examples of how the SPS has been used so far. This demonstrates the kind of outputs the SPS can produce, and the types of investigations that it can be used to perform. The usage permissions and availability of the SPS are explained in Section 3, and Section 4 describes the future development plans for the SPS post-launch. Finally, Section 6 reviews the main conclusions of this paper.
%________________________________________________________________________________________________________________________________

\section{The SPIRE photometer simulator}
The SPIRE photometer simulator (SPS) is a software package which simulates the operation of the SPIRE photometer instrument and associated spacecraft functions. The data are output as time-lines in engineering units, which corresponds to the input data level (Level-0) for the Herschel data pipeline. 

The SPS allows users to input their planned observations, along with representative maps of the sky (either synthetic or derived from existing data), and be returned data which can be reduced and analysed using the official Herschel data pipeline, or any other suitable data reduction software the user wishes to use.  

It is important to note that the SPS should not be used as an observation planning tool as it does not take into account instrument and spacecraft overheads.  All observation planning should be carried out using the Herschel observation planning tool HSpot \citep{hspot}. The simulated data are intended to be a faithful representation of what in-flight data will be like, but while every effort has been made to make the data realistic, it is inevitable that as yet unknown instrument systematics will be present in the post-launch data which are not currently present in the simulated data.    

The SPS is controlled via a graphical user interface (GUI). Observations are set up using a window which contains the same parameters as those in HSpot. A simulated observation can then be performed using any of the SPIRE astronomical observing modes, including parallel mode (only SPIRE output is returned by a parallel mode simulation).

\subsection{Simulator architecture and operation}
The SPS is a modular software package written in the Interactive Data Language (IDL), with each module being controlled by a core program. This design was chosen to allow various modules to be developed independently based on a defined set of module input and output parameters. With the exception of the core program, each module contains four different phases of operation: an \emph{input}, \emph{initialisation}, \emph{temporal}, and \emph{afterrun} phase. The \emph{core} module contains only the \emph{input} phase operation.

The \emph{input} phase loads all of the user defined inputs and stores them in the appropriate data structures in preparation for the simulation run.  The \emph{initialisation} phase uses these input data to compute a variety of standard parameters, including optical transmissions, detector array layout etc. These are parameters which will not change throughout the simulations, and thus can be computed at the start of the simulation. Parameters whose values might change as a result of feedback from other system components are computed in the \emph{step} phase. Most commonly these will be sets of parameters which are mutually dependent. In most cases the majority of a module's operation can be performed in the \emph{initalisation} phase, with the \emph{step} phase allowing for flexibility in the system for future development. Finally, the \emph{afterrun} phase is used to compute any parameters which require entire data time-lines to be complete before their operation can be performed. For example, the 1/\emph{f} noise in the SPS is computed using  Fourier methods, and cannot be computed in a step by step method. Likewise the simulated on-board filtering of the data must be done on entire data time-lines.

\begin{figure}
\centering
\includegraphics[width=8cm]{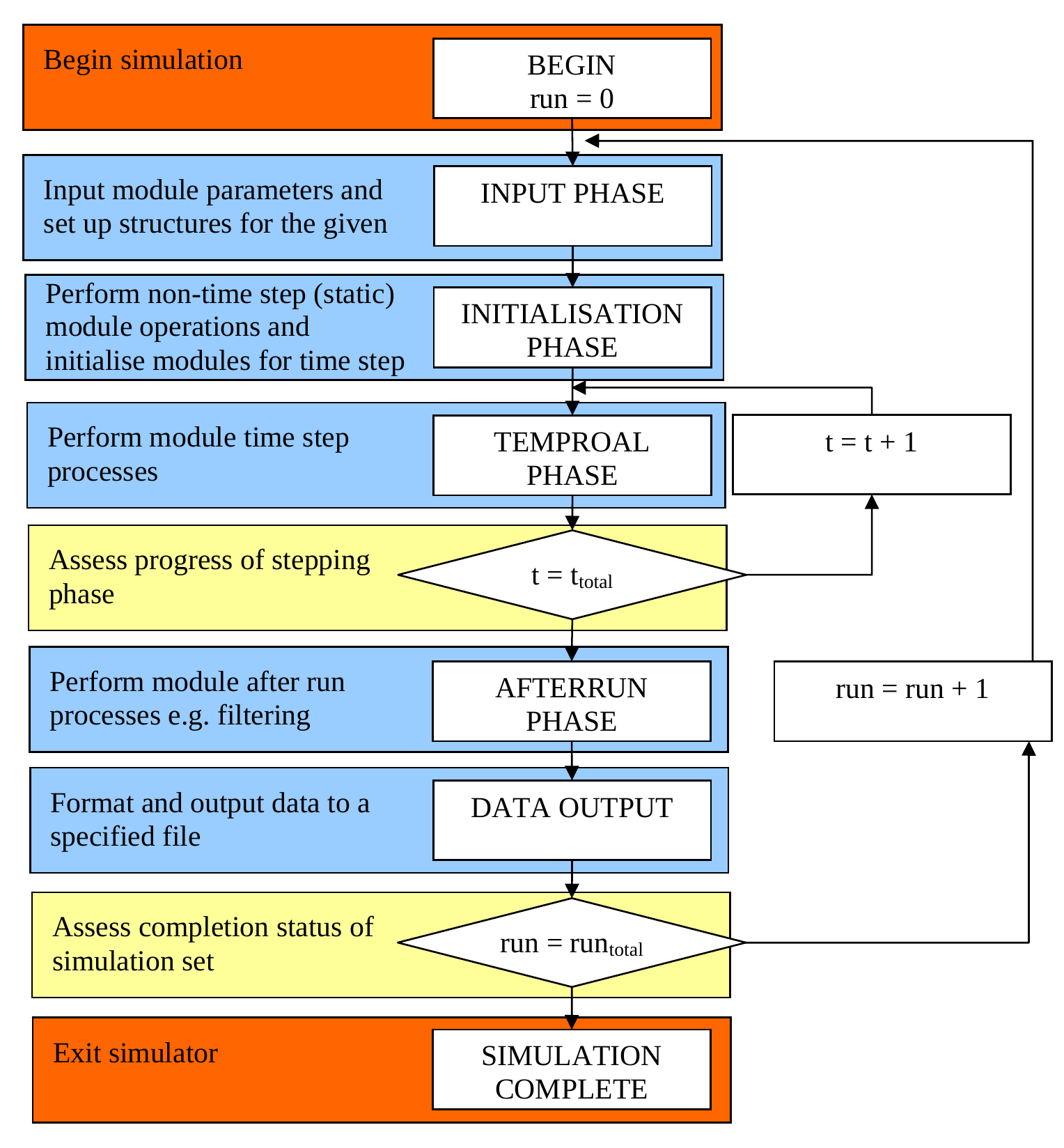}
\caption{Block diagram showing the operational architecture implemented in the SPS.}
\label{block_diagram}
\end{figure}

The processing sequence of the SPS is shown in Figure~\ref{block_diagram} as a flow diagram, with each phase of operation being displayed. This diagram also shows how multiple simulation runs are performed.

\subsection{Modules}\label{modules-section}
The SPS contains 11 modules, with each module representing the operation of a specific system component. For example, the \emph{detectors} module contains all code associated with simulating the detectors, including the generation of noise contributions arising from the detector itself, and the associated load resistor. The modules are named according to the subsystem or function that they simulate and are described in detail below.

\subsubsection{\emph{Core}}
This module does not represent any physical part of the Herschel-SPIRE system. Instead it manages and runs all operations specific to operating the simulation software, such as calling modules in the correct order, performing the required number of simulations, and setting the simulation output file names. It also controls the time step resolution used within the simulator, which directly relates to the simulation run time.

The nominal time step used in the SPS is 21 ms, which equates to 48 Hz, approximately 3 times the nominal instrument detector sampling rate. The internal SPS clock must run faster than the instrument sampling rate to account for short time-scale transient systematics, such as high frequency noise fluctuations. The 3:1 ratio of instrument to internal clock sampling rate was found to give provide an accurate simulated output, with no improvement found when operating at higher ratios.

\subsubsection{\emph{Sky}}\label{input_sky_files}
Like the \emph{core}, this module does not correspond to a particular instrument system. Instead it represents the astronomical sky viewable by the telescope. The module reads in three user defined input sky files, one for each waveband in a Flexible Image Transport System (FITS) file format. The units of the input maps are Jy/pixel, and the pixels must have the same integer arcsecond size for all three bands. It is recommended that a pixel size of 2 arcseconds is used ($\sim$1/10 of the 250 $\mu$m beam). Pixel sizes larger than this can result in digitisation artefacts from the input map being present in the output data. The source spectrum in each pixel is treated uniform across the SPIRE band, therefore any spectral slope information must be included in the input model. This input will, however, be multiplied with the SPIRE filter response function. 

The astrometric data associated with the input images are also read and stored for use in the observatory function module. The astrometic information contained within the FITS header must therefore be correct and contain the CDELT, CRPIX, CTYPE, CRVAL, CROT and EQUINOX keywords.

\subsubsection{\emph{Optics}}
This module represents the main optical properties and parameters of the telescope and the SPIRE photometer (including the filters), and the positional mapping of the detectors on the sky. These parameters are used to derive the optical transmission and emissivity profiles for each optical element, as well as the final instrument spectral response function in each band. This spectral response function is then applied to the input sky model assuming a flat source spectrum across the SPIRE bands. The true source spectrum must be taken in to account in the generation of the input sky model (see Section~\ref{input_sky_files}).

The telescope beam is also taken into account in this module, and is used to produced a beam convolved version of the user's input map. A single indentical beam model is used for each detector within a single array, and three beam models are used in total, one for each band. The beams used are based on optical models and include the telescope side-lobes. The beam model is contained within a library file and loaded from memory when required. Each beam is symmetric and is contained within a square grid of size $\sim$4$\times$4 FWHM.

\subsubsection{Observatory Function}
The observatory function module specifies the photometer observing mode to be simulated in terms of the appropriate observing function and its associated parameters. Observing parameters are supplied via either a simple HSpot style interface, or an advanced interface option. The HSpot option is the nominal input method, and provides input parameters akin to those found in the HSpot software. This does not mean however that the software can read in HSpot output files, rather the input parameters are defined in the same way as those in HSpot. The advanced screen allows the user to select different options in order to investigate the effect of different observing strategies which are not normally permitted by Herschel. 

The user input information supplied is then used to generate both the telescope pointing information, including pointing errors, as well as the input parameters for the \emph{BSM} module.

\subsubsection{\emph{BSM}}
This module simulates the movement of the BSM. It generates pointing time-lines in terms of the fixed array spacecraft axes. These time-lines are generated originally in arcseconds, and then converted to analog-to-digital units (ADU) via a look-up table. Both outputs are provided to the user.

\subsubsection{\emph{Thermal}}
This module simulates all information pertaining to the temperatures of the instrument and the telescope, and their temporal fluctuations. However, since the instrument system, and telescope can be considered to be stable on time scales of a single observation, the model used in the SPS assumes that the only source of thermal fluctuations is the $^{3}$He cooler. The cooler fluctuations are simulated using a simple 1/\emph{f}$^{\alpha}$ noise time-line generator, with a user specified noise spectral density, knee frequency, and spectral index ($\alpha$).

The cooler fluctuation time-line is low-pass filtered assuming an RC filter profile. This represents the suppression of high frequency thermal variations by the various thermal linkages between the cooler tip and the array, and the thermal mass of the detector arrays themselves. A different user defined time constant is used for each array, representing the different distances and linkages between the cooler tip and each specific array. This results in the three arrays having a similar, although not identical thermal fluctuation time-line. A demonstration of the impact of thermal drifts on the output timeline can be seen in Figure~\ref{sample_timeline}(a), along with the reference astronomical power timeline.

At present there are no thermal gradients within a bolometer array, meaning that all bolometers vary in temperature simultaneously. These thermal variations represent the dominant source of common mode 1/\emph{f} noise in SPIRE, and as with all systematics simulated, can be turned on and off depending on the users needs.

\subsubsection{\emph{Astronomical Power}}
This module uses both the telescope and BSM pointing information to generate absorbed power time-lines for each detector. The observing strategy is projected onto the beam convolved input sky, and the flux from each pointing is registered. Additional telescope efficiency factors and filter transmission profiles are then used to compute the astronomical power falling on a given bolometer during each simulation time step. A sample output timeline is given by the dashed line in Figure~\ref{sample_timeline}.

\subsubsection{\emph{Background Power}}	
The background power falling on each bolometer throughout a simulation is generated in this module. Each element in the optical chain is assumed to emit as a blackbody, with a wavelength dependent emissivity of $(1-t)$, where $t$ is the optical transmission. The primary mirror is modelled with a user defined thermal gradient (with temperature decreasing with distance from the spacecraft's sun shield). The background power of the Herschel/SPIRE system is expected to be stable on the period of a single observation, therefore no background power variations are implemented. This will be reassessed once post-launch data are available. 

\subsubsection{\emph{Detectors}}
This module produces an output voltage time-line for each detector channel based on inputs from the \emph{Astronomical} and \emph{Background power} modules. The bolometer thermal model presented by \cite{Sudiwala} and \cite{Woodcraft2008} is used to calculate the correct voltage output based on the power absorbed by the detector.  The thermal model provides an accurate characterisation of non-linear bolometer response to high levels of incident power, and detector output fluctuations due to variations in the bath temperature.  Each detector is modelled independently using its own set of bolometer parameters measured during instrument testing.

In addition to the detector voltage time-lines, the theoretical photon, Johnson, phonon and amplifier noise contributions are all calculated here. These data are then used to generate unique white and/or 1/$f$ noise time-lines, which are in turn summed with the detector output to produce the final detector voltage time-line (see Figure~\ref{sample_timeline}(b)).

\subsubsection{\emph{Sampling}}
The \emph{sampling} module simulates the processing of the output detector time-lines by the on-board read-out electronics. Each detector time-line is filtered by the on-board 5-Hz low-pass filter, digitized, and output at the requested sampling frequency. A voltage offset is also removed from the time-lines, and output as a separate single value for each detector.

\begin{figure}
\centering
\includegraphics[width=9cm]{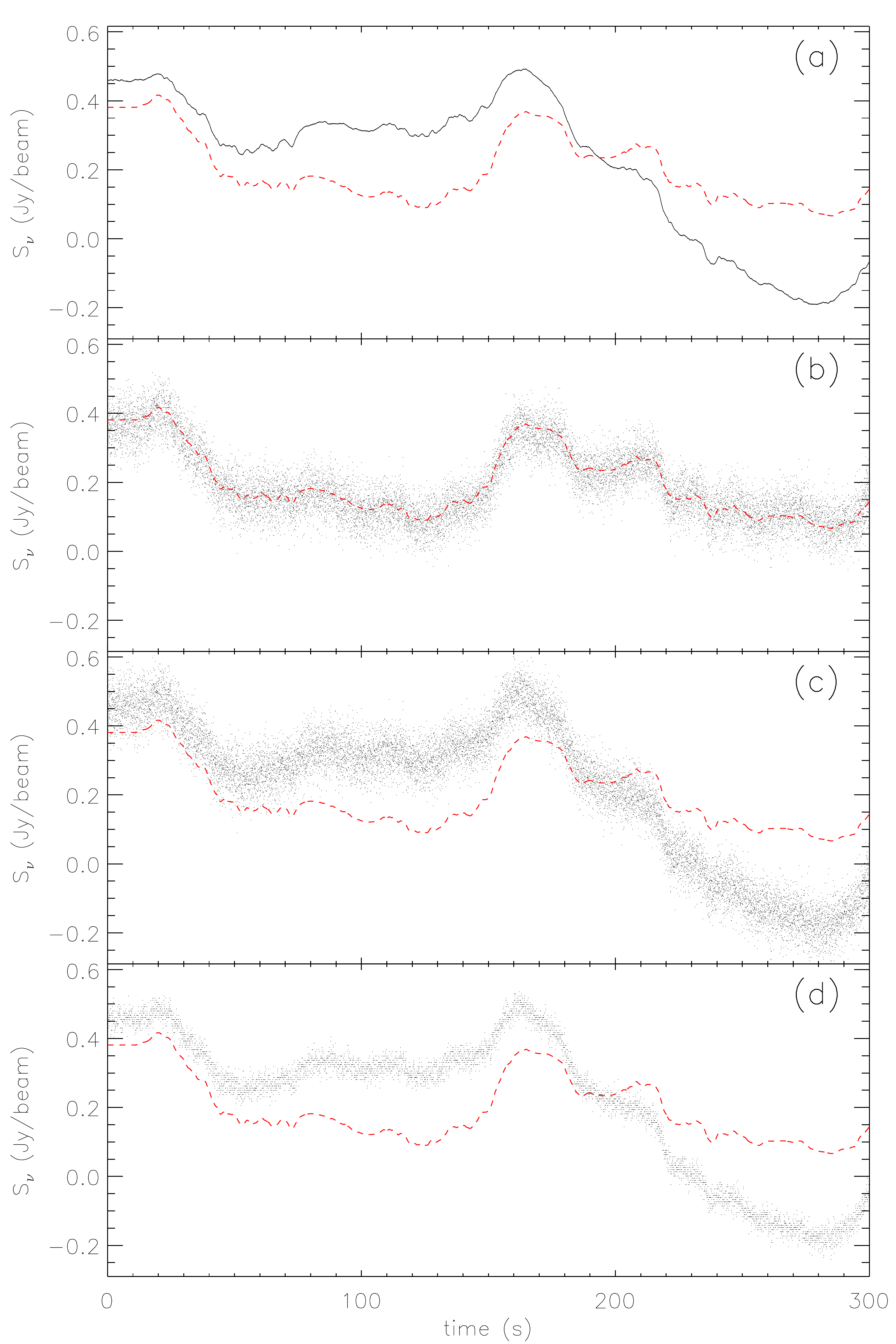}
\caption{Sample data timeline for the central bolometer in the 250~$\mu$m array under various noise conditions: (a) contains only thermal drifts (correlated 1/\emph{f} noise) with no other noise effects; (b) shows the timeline with uncorrelated 1/\emph{f} and white noise applied; (c) contains both of the previous two outputs combined, and (d) shows the timeline from (c) once it has passed through the read-out electronics, i.e. low-pass filtered, sampled, and digitised. All signals have been calibrated to Jy/beam from their native unit to allow direct comparison for direct comparison of the different noise effect. The original noiseless astronomical power timeline is also given in each panel by the dashed line.}
\label{sample_timeline}
\end{figure}

Sample input and output time-lines to this module are given in Figures~\ref{sample_timeline}(c) and (d) respectively. The decreased noise level resulting from the low-pass filtering, along with the digitisation of the signal are evident in these panels.

\subsubsection{\emph{Housekeeping}}
The \emph{housekeeping} module outputs various instrument monitoring information. These data include flags which identify the observing mode being used, and the building blocks used to make up that mode.

\subsection{Output Data}
The SPS output is a standard FITS file with 11 extensions. Each extension contains the data structure from one of the SPS modules. All input parameters are contained within the output file, however for file size reasons not all derived time-lines are stored as standard. Nominally, only the time-lines returned by the physical system are output, i.e. telescope pointing and digitised bolometer outputs. It is possible, however, to choose to output the astronomical power, background power, noiseless voltage, and noisy voltage time-lines if requested. These time-lines are sampled at a higher frequency, and hence can significantly increase the output file size.

The standard FITS output can be loaded into the official Herschel pipeline processing software via a separate piece of converter software, whereupon it can then be reduced in the same way as real SPIRE data. Alternatively the standard output can be converted from digitized units (level-0) to calibrated data (level-1) time-lines using a conversion routine supplied with the SPS. This new file can then either be loaded into the Herschel pipeline, or read in using another piece of custom software.

\subsection{System requirements, performance and limitations}
The SPS will run on any modern computer (CPU $\sim$2 GHz, RAM $\sim$2 GB) and requires IDL V6.2 or higher. Using a typical computer the SPS will run approximately 4 times faster than real time, meaning it would take 1 hour to simulate a 4 hour observation. Typically a simulated one square degree scan map observation will require $\sim$635 MB of memory during the simulation, and generate a $\sim$44 MB output file. The memory required during a simulation, and the size of the output file are both dependent on the size of the input file, and the selection of output parameters. In an optimal configuration for specific investigations the system has been shown to run up to 10 times faster than real time. Simulated observations can be performed for the full 18 hours available in a single Herschel astronomical observing request.

A lack of detail in the spacecraft pointing model means that SPS cannot be used to calculate observing times. Many of the required overheads such as slewing and pointing calibrations are not included in the SPS. The pointing model used has been designed to provide realistic data for on-source regions only.

The SPS contains all of the major known instrument systematics, and represents the best current estimate of the SPIRE photometer's in-flight performance.

%________________________________________________________________________________________________________________________________
\section{Uses of the SPS}
The SPS has been used for a variety of purposes throughout its development.  It was used to optimise and characterise the SPIRE photometer astronomical observing modes, and identify the best observing strategy possible with SPIRE \citep{Sibthorpe2006, Waskett2007, Sibthorpe2008}. It has also provided input data for a variety of map-making algorithms during a map-making selection exercise. The different algorithms were compared, and the best chosen as the standard map-making algorithm for use in the Herschel pipeline (Chanial, submitted to PASP) 

In addition to mission planning, the SPS is currently being used in the preparation and planning of various Herschel key programmes, both in guaranteed and open time. These programmes span a very wide range of targets, from high redshift galaxy surveys through to observations of diffuse galactic structure and star forming regions.  Investigation of saturation in surveys containing high dynamic range, uniformity of noise in output maps, and development and testing of specific algorithms for use with survey data are all examples of recent uses of the SPS.

\begin{figure}
\centering
\subfigure[\label{hi_z_map   }]{\includegraphics[width=8cm]{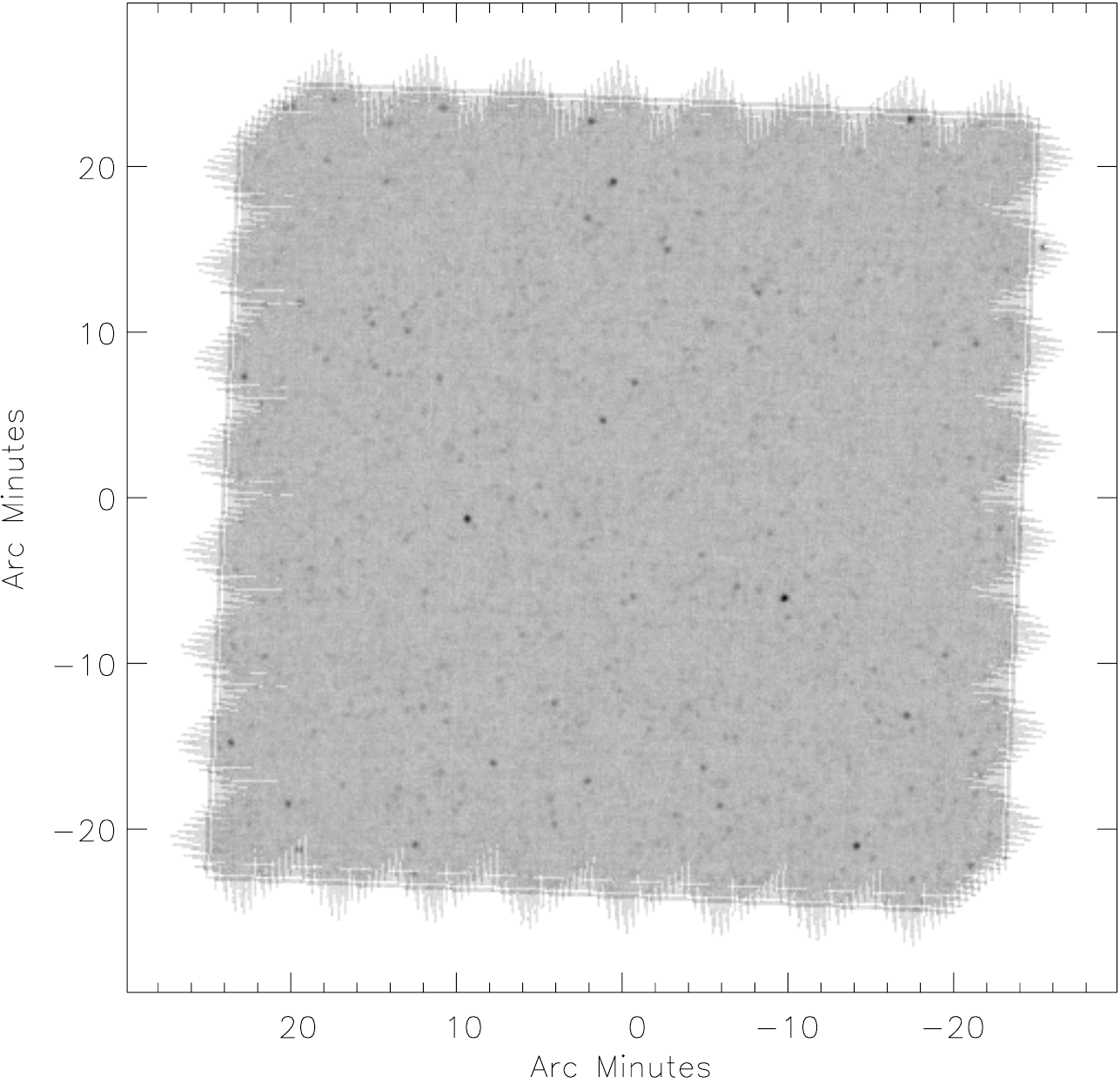}}
\subfigure[\label{hi_z_counts}]{\includegraphics[width=8cm]{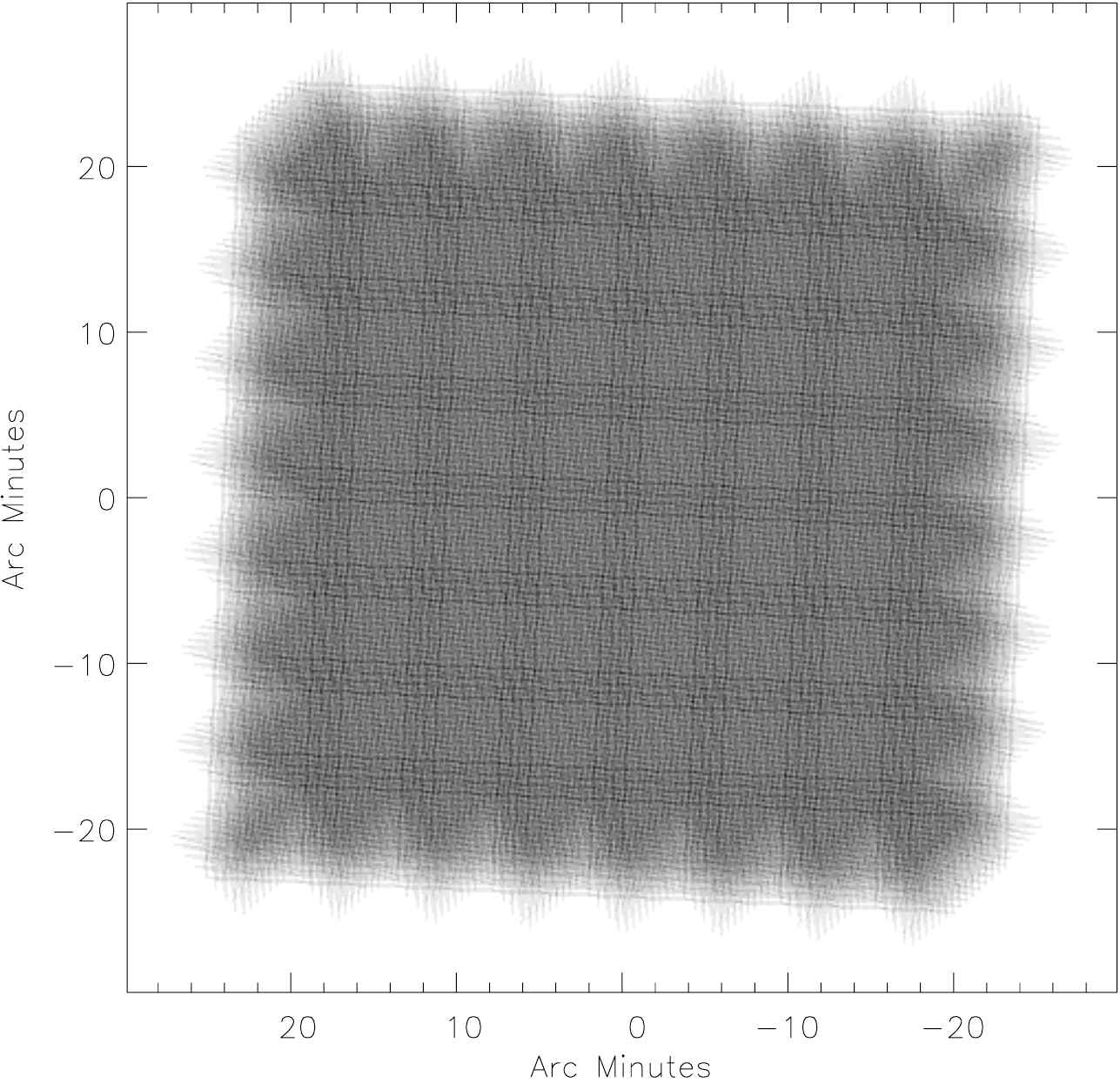}}
\caption{Simulated `large map' observation of a high redshift extra-galactic field. (a) This image is from the 250 $\mu m$ band, and shows an image of size 45 arcmin $\times$~45 arcmin. A na\"ive map-maker was used to construct the image from the output time-line data. (b) coverage map showing uniformity of the observing time.}
\label{hi_z_sim}
\end{figure}

An example output produced during planning for one of the high redshift galaxy surveys is presented in Figure~\ref{hi_z_sim}. The input to this simulation is a map derived from galaxy catalogues which were scaled to the appropriate flux densities. Three input images were created, one for each band, but only the 250 $\mu$m band image is shown here. Two orthogonal simulated `large map' scans were performed with 1/$f$ noise included. The time-line data from these two observations were then made into a na\"ive map and co-added to produce the final image seen in Figure~\ref{hi_z_sim}a. This type of image can be used to investigate the kind of problems this type of survey might have with source extraction, P(D) analysis, and various other data analysis techniques. A coverage map, also output by the map-maker (Figure~\ref{hi_z_sim}b), provides additional information on the uniformity of integration time, and hence depth across the map. With these two maps the noise statistics of the data can be investigated, and any biases characterised. With a known input sky model, methods which attempt to remove the various sources of noise can also be tested and compared.

\begin{figure}
\centering
\subfigure[\label{debris_in }]{\includegraphics[width=4cm]{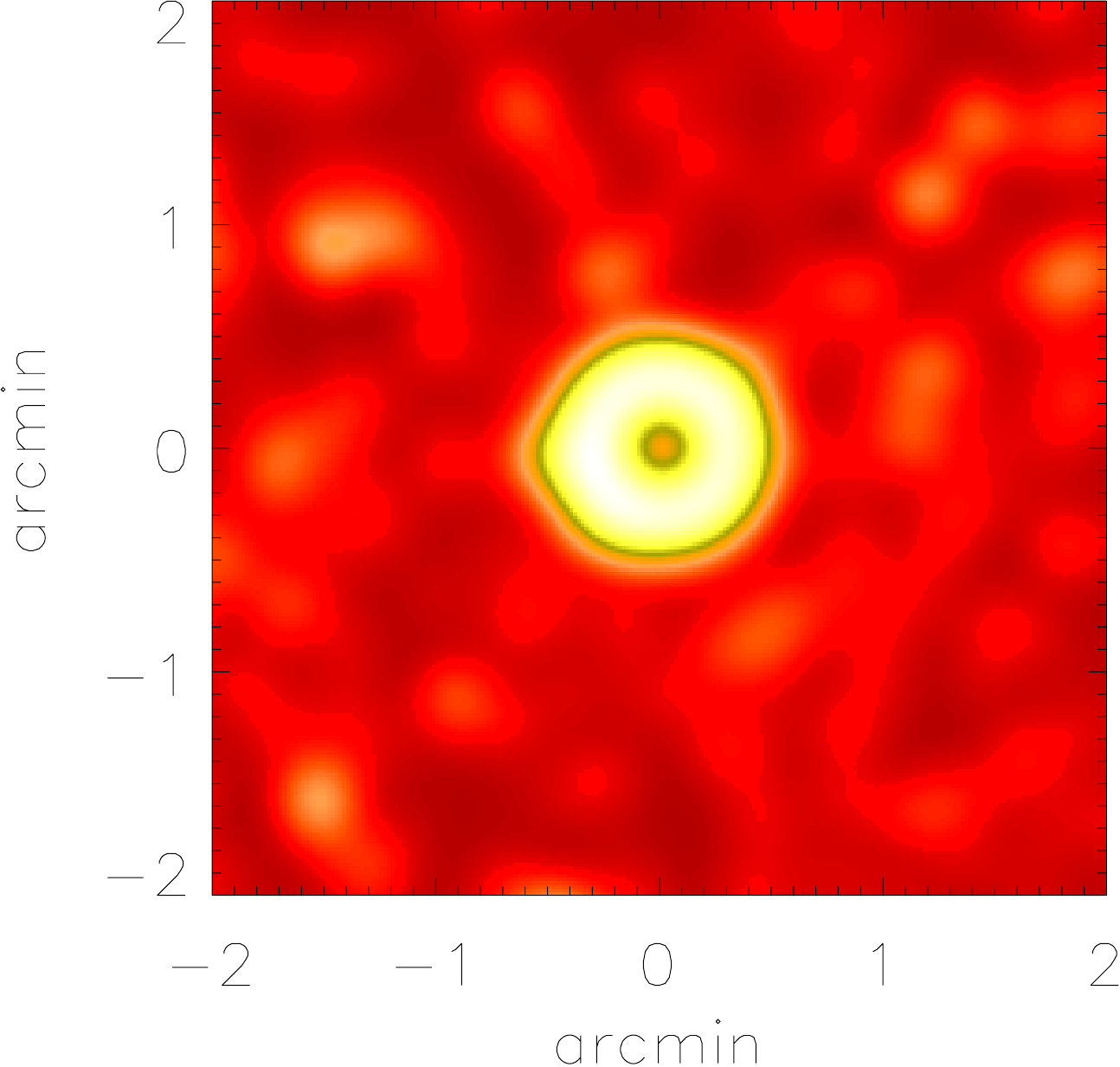}}
\subfigure[\label{debris_jig}]{\includegraphics[width=4cm]{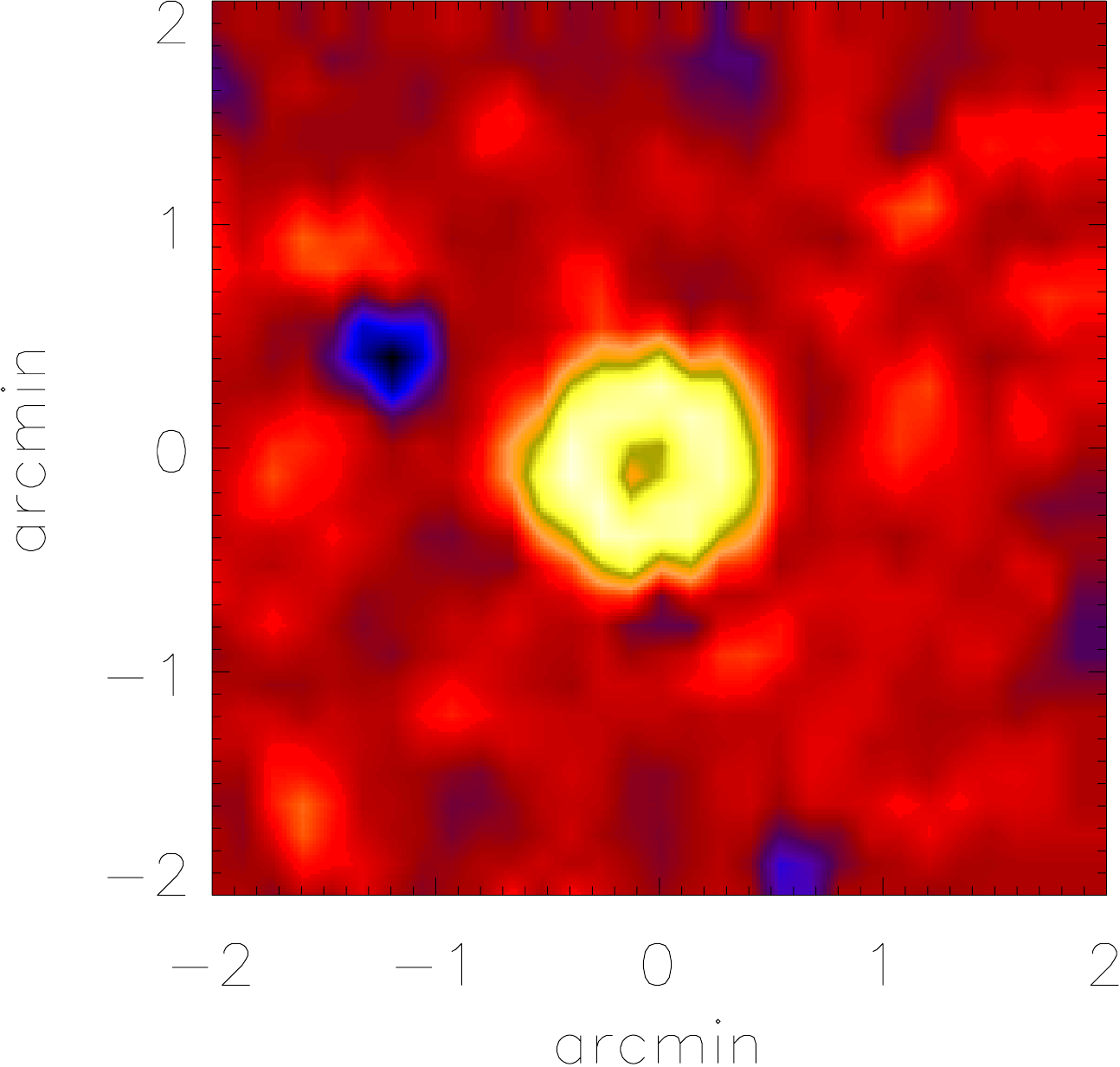}}
\caption{Noiseless simulated `small map' observation of a debris disc -- (a) input sky model, (b) output map.}
\label{debris_sim}
\end{figure}

An example of a `small map' observation is given in Figure~\ref{debris_sim}. Here the observation of a debris disc, whose brightness distribution is based on that measured for the Epsilon Eridani disc \citep{Greaves1998}, has been simulated without instrumental noise. The input sky model is shown in Figure~\ref{debris_sim}a, and consists of a simple torus shaped object placed on top of a realistic extra-galactic background. The goal of this simulation was to provide a way to study the influence of extra-galactic confusion on the derived properties and structure of the debris disc when performing a chopped SPIRE observation. This investigation found that a `small map' observation results in a $\sim$20\% higher extra-galactic confusion noise level than the same observation performed in `large map' mode \citep{Sibthorpe_spitzer_conf_2008}. In addition, it was found that the structure of the background galaxies in the `small map' simulation was far more uncertain as a result of their emission being mixed in the output with that of sources in the off-source field. Consequently, methods to identify and remove the contaminating sources are significantly more difficult to employ.

The information from these types of simulations allows for a better informed choice of observing mode. A mode can be selected which fits the type of source, and confusion noise environment. Simulations can also be used to develop, test, and characterise a series of analysis routines so that they are ready once the data are taken.
%________________________________________________________________________________________________________________________________
\section{Usage and availability}
The SPS software is publicly available from the SPS website (\url{www.roe.ac.uk/~sib/sps}), along with the required conversion software which enables the SPS output to be used with HIPE. All relevant information on how to obtain and use the SPS can be found at on the SPS website, along with further details on the models used within the SPS. This site will also contain information on updates to the SPS as they are released\footnotemark[1]{}.

\footnotetext[1]{The SPS may be used free of charge only for non-commercial research purposes. If you make modifications to this software, you must clearly mark the software as having been changed and you must also retain the copyright and disclaimer. In the interests of avoiding the propagation of modified versions of the SPS, we would request that you do not edit or redistribute this software without first contacting the author, or a member of the SPS team. All uses of this software must acknowledge it's use, and reference this paper.}
%________________________________________________________________________________________________________________________________
\section{Future development}
Following the launch of Herschel a new version of the SPS will be released. This version will be updated to include in-flight performance data, and will aim to replicate the true characteristics of the operating system. Feedback from users is also expected to guide the development of additional features in the SPS.
%________________________________________________________________________________________________________________________________
\section{Conclusions}
In this paper we have described the SPIRE photometer simulator software and given examples of its capabilities and outputs. The SPS mimics the output of the SPIRE photometer such that it can be loaded into the Herschel pipeline environment, and reduced as if it were real data. The software represents the primary instrumental characteristics, and allows the user to investigate how these characteristics might influence the quality of astronomical data for individual science cases. It also provides an opportunity to prepare, and characterise, science case specific data analysis routines.

We have presented several examples of preparation work carried out with the SPS, and provided information on how new users can access and use the SPS in preparation for their own Herschel observations. The SPIRE photometer simulator is a versatile tool for the investigation of a wide range of instrumental influences on astronomical data obtained with Herschel-SPIRE.
%   \begin{enumerate}
%   \end{enumerate}

\begin{acknowledgements}
The authors wish to acknowledge significant contributions to the SPIRE simulator project from Pierr\'e Chanial, Trevor Fulton, Ren\'e Gastaud, Michael Pohlen, Emma Rigby, Richard Savage, Rupert Ward, Tim Waskett, Steven Watkin, and Adam Woodcraft. We also acknowledge the Science and Technology Facilities Council for post-doctoral funding of this work.
\end{acknowledgements}
%\begin{thebibliography}{}
%\end{thebibliography}
\bibliographystyle{aa}
\bibliography{references}

\end{document}